\documentstyle[sprocl,epsfig]{article}

\def\be{\begin{equation}}
\def\ee{\end{equation}}
\def\bea{\begin{eqnarray}}
\def\eea{\end{eqnarray}}

\begin{document}
\title{
SEARCH FOR EXTRA GAUGE BOSONS IN LITTLE HIGGS MODELS 
AT A LINEAR COLLIDER
\\}

\author{ GI-CHOL CHO }

\address{
Department of Physics, Ochanomizu University,\\
Tokyo, 112-8610, Japan
}

\author{AYA OMOTE}

\address{
Graduate School of Humanities and Sciences, \\
Ochanomizu University, Tokyo, 112-8610, Japan
}

\maketitle\abstracts{
A generic feature of little Higgs models is presence of extra neutral
gauge bosons. In the littlest Higgs model, the neutral extra gauge boson 
$A_H$ is lightest among the extra particles and could be as light as a 
few hundred GeV, which may be produced directly at an $e^+ e^-$ linear 
collider. 
We study production and decay of $A_H$ at the linear collider and
compare them with those of $Z'$ bosons in supersymmetric $E_6$
models. 
%
}

\section{Introduction}
The idea of little 
Higgs~\cite{Arkani-Hamed:2001nc,Arkani-Hamed:2002pa,Arkani-Hamed:2002qx} 
has been proposed as an alternative of supersymmetry to solve the gauge 
hierarchy problem. 
In this class of models, the electroweak Higgs boson appears as 
a pseudo-goldstone boson of a certain global symmetry breaking 
at a scale $\Lambda \sim 10~{\rm TeV}$ 
so that the Higgs boson mass can be as light as $O(100~{\rm GeV})$. 
The light Higgs boson mass is protected from the 1-loop quadratic
divergence by gauging a part of global symmetry, and introducing 
a few extra heavy particles whose typical mass scale is of order 
$f \equiv \Lambda/4\pi$, where $f$ is a decay constant of the 
pseudo-goldstone boson. 
A generic feature of the little Higgs models is a larger gauge symmetry 
than the Standard Model (SM), which is broken near the electroweak
scale, so that there are some extra gauge bosons at the TeV scale. 
%

%
In this talk, we would like to study production and decay of the extra 
gauge bosons at a future $e^+ e^-$ linear collider.  
Especially we focus on the Littlest Higgs Model (LHM) which is the 
simplest and earliest one in this class of 
models~\cite{Arkani-Hamed:2002qy}. 
The LHM has $[{\rm SU(2)}\times {\rm U(1)}]^2$ symmetry at ultra-violet
regime, and which is broken to that of the SM, SU(2)$_L\times$U(1)$_Y$, 
at the scale $\Lambda$. 
Then, there are four massive extra gauge bosons and they are mixed with
the SM gauge bosons after the electroweak symmetry breaking. 
As a result, the set of extra gauge bosons at the weak scale consists of
electrically neutral states ($A_H, Z_H$) and charged states 
($W_H^\pm$). 
Among them, the $A_H$ boson is lightest so that it is expected to be 
discovered at future collider experiments rather early. 
The $A_H$ boson in hadron collider experiments has been studied in 
refs.~\cite{Han:2003wu,Hewett:2002px}. 
%

%
Since such an experimental signature of the $A_H$ boson is quite similar 
to a $Z'$ boson, it is very important to identify the models if an extra 
neutral gauge boson is discovered in hadron collider experiments. 
Although the LHC experiment could discover the $Z'$ boson up to 
3 TeV~\cite{Golutvin:2003gc}, it is hard to test the $Z'$ models. 
On the other hand, an $e^+ e^-$ linear collider can play a 
complementary role for such purpose~\cite{Aguilar-Saavedra:2001rg}. 
In this talk, we compare the experimental signatures of $A_H$ and  
those of $Z'$ boson in supersymmetric $E_6$ models~\cite{Hewett:1988xc}, 
and examine a possibility to distinguish these models in the linear 
collider experiments. 
In our study, we assume that Tevatron or LHC discovers a certain $Z'$ 
boson whose mass is smaller than $\sqrt{s}$ of the linear collider. 
Then we can study the $Z'$ boson at the linear collider by tuning 
the $e^+ e^-$ beam energy at the peak of $Z'$ resonance. 
We also assume that the mixing with the SM $Z$ boson is negligibly small 
because such a mixing is severely constrained from the experimental data 
of the electroweak precision measurements at the
$Z$-pole~\cite{Cho:1998nr}. 
%

%
The interaction of $A_H$ to a fermion $f$ in the SM is described by the 
following Lagrangian~\cite{Han:2003wu}: 
\begin{eqnarray}
{\cal L} &=& -\frac{g_Y}{2 \sin\theta' \cos\theta'}
Q_{f_\alpha}^{A_H} \overline{f_\alpha} \gamma^\mu f_\alpha 
A_{H\mu}, 
\label{interaction}
\end{eqnarray}
where $\alpha(=L,R)$ denotes the chirality of fermion $f$, and 
$\theta'$ is the mixing angle of two U(1) gauge bosons. 
The charge $Q^{A_H}_{f_\alpha}$ for the fermion $f_\alpha$ 
is summarized in ref.~\cite{Cho:2004xt}. 
The interactions and couplings of $Z'$ to the fermion pairs in the SUSY 
$E_6$ models -- $\chi,\psi,\eta$ and $\nu$ models -- 
can also be found in ref.~\cite{Cho:2004xt}. 
%

\begin{figure}[ht]
     \begin{center}
     \begin{tabular}{cc}
     \mbox{\epsfig{file=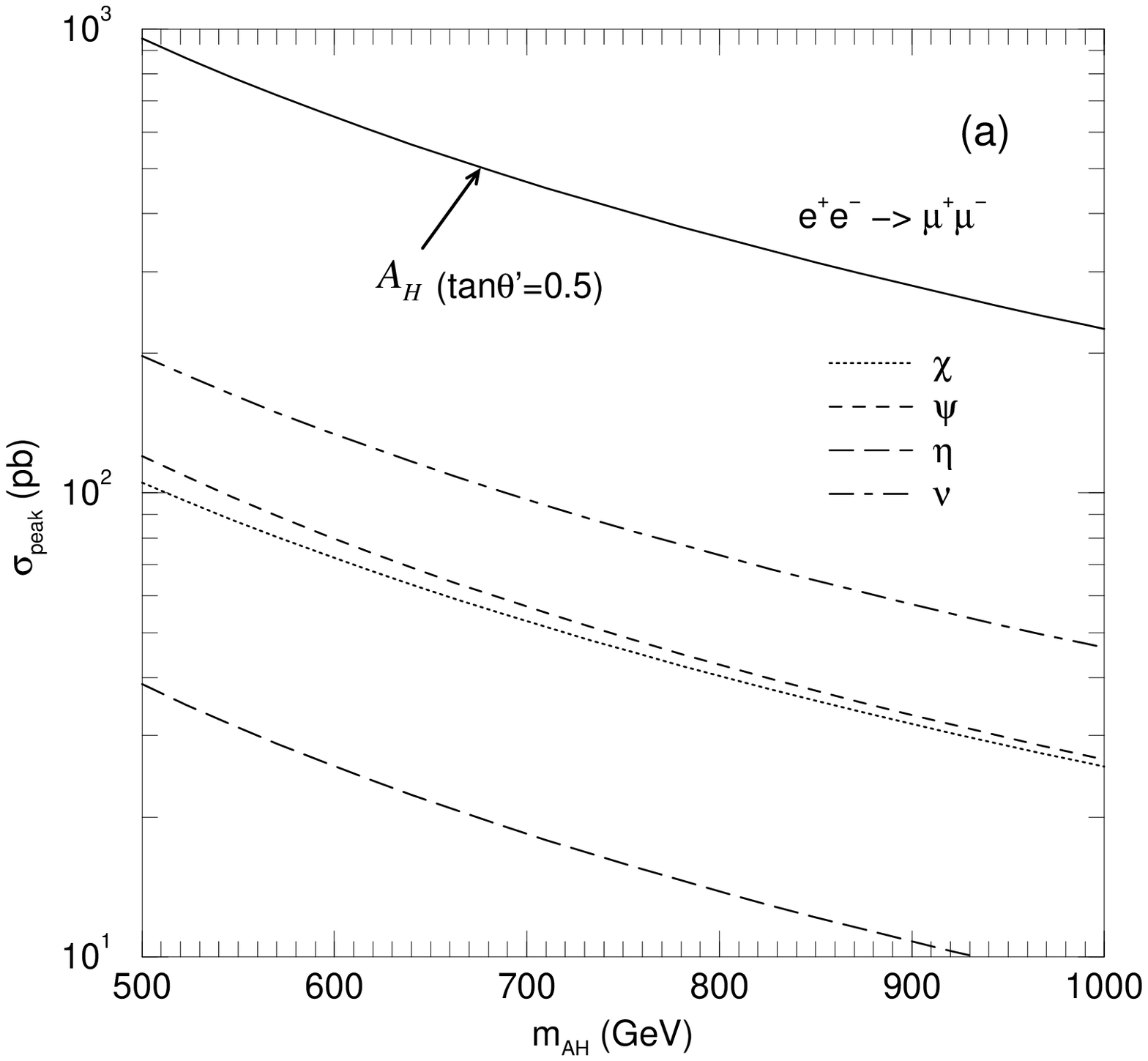,width=5.7cm}}&
     \mbox{\epsfig{file=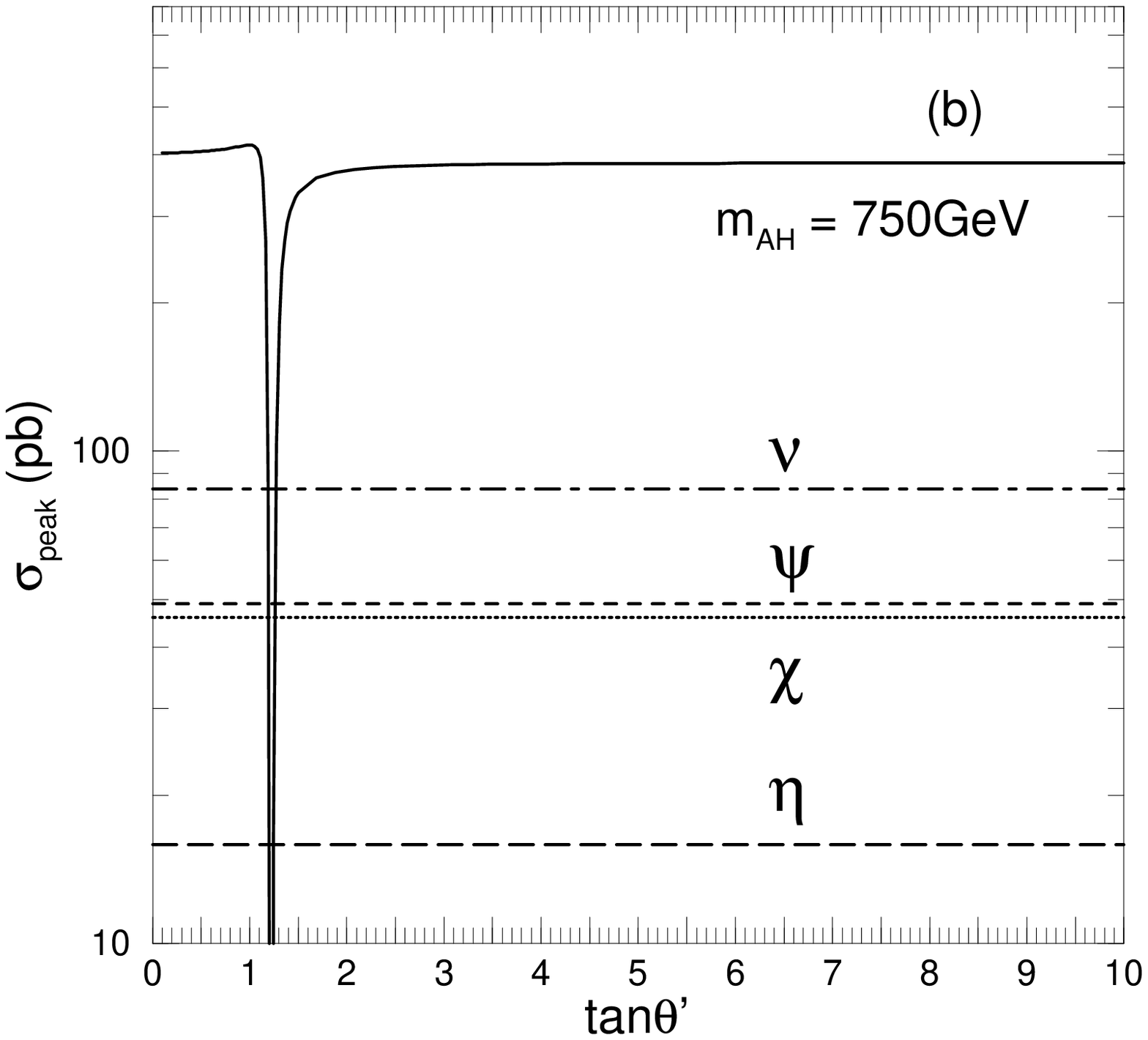,width=5.7cm}}
     \end{tabular}
     \end{center}
\caption{The peak cross section of $e^+ e^- \to \mu^+\mu^-$ at the $Z'$ 
pole in the LHM (solide line) and SUSY $E_6$ models as functions of 
$A_H$ (or $Z'$) mass (a), and of the mixing angle $\theta'$ (b). 
}
\label{fig:peak}
     \end{figure}

%
We show the peak cross section of $e^+e^- \to \mu^+\mu^-$ in the 
LHM and SUSY $E_6$ models as a function of the $A_H(Z')$ mass in 
Fig.~\ref{fig:peak}(a), and of the mixing angle $\tan\theta'$ in 
Fig.~\ref{fig:peak}(b). 
In Fig.~\ref{fig:peak}(a), the LHM prediction is shown for 
$\tan\theta'=0.5$. 
On the other hand, Fig.~\ref{fig:peak}(b) is obtained for 
$m_{A_H}(m_{Z'})=750~{\rm GeV}$. 
We can see in Fig.~\ref{fig:peak}(a) that the peak cross section in 
the LHM is roughly a few hundred pb, which is a few times larger 
than those in SUSY $E_6$ models, 
so the cross section measurement seems to be useful to test the models 
at the linear collider with $100 {\rm fb}^{-1}$ integrated luminosity.  
However, we can see in Fig.~\ref{fig:peak}(b) that the peak cross 
section in the LHM rapidly decreases around $\tan\theta' = 1.2$. 
This is because that both the left- and right-handed electron couplings
to $A_H$ are proportional to $-\frac{2}{5}+\cos\theta'^2$, 
and they diminish for $\tan\theta'\sim 1.2$. 
Therefore, we should find another observable to test the models. 
%

%
The forward-backward (FB) asymmetry of the $e^+ e^- \to f \bar{f}$ 
process does not have the $\theta'$ dependence. 
The FB asymmetry at the pole of $A_H(Z')$ can be expressed as 
\begin{eqnarray}
A_{\rm FB}^f &=& \frac{3}{4} A^e A^f, 
\label{eq_asym}
\end{eqnarray}
and $A^f$ can be expressed in terms of the parameter $r_f$ which is 
defined as a ratio of the left- and right-handed 
couplings~\cite{Cho:2004xt}: 
\begin{eqnarray}
A^f = \frac{1-r_f^2}{1+r_f^2}. 
\end{eqnarray}
It is easy to see that the parameter $r_f~(f=e,u,d)$ in the LHM is 
independent of $\theta'$; 
\begin{eqnarray}
(r_e,r_u,r_d)=(2,4,-2). 
\end{eqnarray}
The FB asymmetry, therefore, is a good observable to compare the 
LHM and the SUSY $E_6$ models. 
The FB asymmetry for the muon, $c$-quark and $b$-quark in the LHM 
and the SUSY $E_6$ models is summarized in Table~\ref{asym}. 
It is remarkable that the asymmetries in the $\psi$ model are zero
because that the extra U(1) charge assignments on the SM fermions are 
parity invariant. 
Beside on the $\psi$-model, the difference of predictions between 
the LHM and the SUSY $E_6$ models are very clear in the $b$- and 
$c$-quark asymmetries. 
In the $b$-quark FB asymmetry, the LHM predicts a positive value while 
the SUSY $E_6$ model are negative one. 
Especially, it is noticeable that there is no $c$-quark asymmetry in 
the SUSY $E_6$ models though the LHM gives a 40\% asymmetry.  
The reason why the FB asymmetry of $c$-quark vanishes in SUSY $E_6$ 
models is as follows. 
As shown in eq.~(\ref{eq_asym}), the FB asymmetry is given by the 
difference of the couplings between the left- and right-handed 
fermions to the $Z'$ boson. 
In SUSY $E_6$ models, both left- and right-handed $c$-quarks are
embedded in the same multiplet, $\mathbf{10}$ representation in SU(5), 
so that they have a common coupling which leads to no asymmetry.  
Table~\ref{asym} tells us that the measurements of $b$- and 
$c$-quark asymmetries in a few \% accuracy is enough to test 
if a $Z'$ boson is $A_H$ in the LHM or one of the SUSY $E_6$ models.  
We, therefore, conclude that the measurements of FB-asymmetries for 
heavy quarks are very useful to test if the $Z'$ boson is $A_H$ in the 
LHM or one of the SUSY $E_6$ models.

\begin{table}
\begin{center}
\begin{tabular}{|c||c|c|c|c|c|} \hline
 & $A_H$ & $\chi$ & $\psi$ & $\eta$ & $\nu$ \\ \hline
$A_{\rm FB}^\mu$ & 0.27 & \hphantom{-} 0.48 & 0 
& \hphantom{-} 0.27 & \hphantom{-} 0.27 \\
$A_{\rm FB}^b$ & 0.27 & $-0.48$ & 0 & $-0.27$ & $-0.27$ \\
$A_{\rm FB}^c$ & 0.40 & 0 & 0 & 0 & 0 \\ \hline
\end{tabular}
\end{center}
\caption{Forward-backward asymmetry $A_{\rm FB}$ for 
muon, $b$ and $c$ quarks in LHM and SUSY $E_6$ models. 
}
\label{asym}
\end{table}


\end{document}